\documentclass{elsart}
\usepackage{epsf,epsfig,amssymb,latexsym}

\newcommand{\figwidth}{0.9\textwidth}
\newcommand{\figwidthb}{0.5\textwidth}

\begin{document}
\begin{frontmatter}

  \title{Scaling the localisation lengths for two interacting
    particles in one-dimensional random potentials}


\author[TUC]{Rudolf A. R\"omer}, 
\author[MPIPKS]{Mark Leadbeater}, and
\author[TUC]{Michael Schreiber}

\address[TUC]{Institut f\"{u}r Physik, Technische Universit\"{a}t,
  D-09107 Chemnitz, Germany}

\address[MPIPKS]{Dipartimento di Fisica, Universita di Roma
  Tre, Via della Vasca Navale 84, I-00146 Roma, Italy}

\date{$Revision: 1.4 $; compiled \today}


\begin{abstract}
  Using a numerical decimation method, we compute the localisation
  length $\lambda_{2}$ for two onsite interacting particles (TIP) in a
  one-dimensional random potential. We show that an interaction $U>0$
  does lead to $\lambda_2(U) > \lambda_2(0)$ for not too large $U$ and
  test the validity of various proposed fit functions for
  $\lambda_2(U)$. Finite-size scaling allows us to obtain infinite
  sample size estimates $\xi_{2}(U)$ and we find that $ \xi_{2}(U)
  \sim \xi_2(0)^{\alpha(U)} $ with $\alpha(U)$ varying between
  $\alpha(0)\approx 1$ and $\alpha(1) \approx 1.5$. We observe that
  all $\xi_2(U)$ data can be made to coalesce onto a single scaling
  curve. We also present results for the problem of TIP in two
  different random potentials corresponding to interacting
  electron-hole pairs.
\end{abstract}


\end{frontmatter}

%
%



In two recent articles \cite{LRS98a,LRS98b}, we studied as a simple
and tractable approach to the problem of interacting electrons in
disordered materials the case of only two interacting particles (TIP)
in 1D random potentials. Previous considerations \cite{S94} had led to
the idea that attractive as well as repulsive interactions between TIP
give rise to the formation of particle pairs whose localisation
length $\lambda_2$ is much larger than the single-particle (SP)
localisation length $\lambda_1\approx 105/W^2$,
\begin{equation}
  \lambda_2 \sim U^2 \lambda_1^2
  \label{eq-shep}
\end{equation}
at two-particle energy $E=0$, with $U$ the Hubbard interaction
strength. Although many papers have numerically investigated the TIP
effect \cite{S94,FMPW95,SK97,RS97,PS97,RS98,HMK98}, an unambiguous
reproduction of Eq.\ (\ref{eq-shep}) is still lacking.  However, it
appears well established that some TIP delocalisation such as
$\lambda_2 > \lambda_1$ does indeed exist due to the interaction.
Recently, a duality in the spectral statistics for $U$ and
$\sqrt{24}/U$ has been proposed \cite{WWP98} for small and very large
$|U|$.

In Refs.\ \cite{LRS98a,LRS98b}, we have employed a numerical
decimation method \cite{LW80}, i.e., we replaced the full Hamiltonian
by an effective Hamiltonian for the doubly-occupied sites only.  In
\cite{LRS98a}, we considered the case of TIP with $n$, $m$
corresponding to the positions of each particle on a chain of length
$M$ and random potentials $\epsilon^1_n =\epsilon^2_n \in [-W/2,W/2]$.
In \cite{LRS98b}, we studied the case where $\epsilon^1_n$ and
$\epsilon^2_n$ are chosen independently from the interval
$[-W/2,W/2]$, which may be viewed as corresponding to an electron and
a hole on the same chain (IEH).
Via a simple inversion, we then obtained the Green function matrix
elements $\langle 1,1\vert {G_2}\vert M,M\rangle$ between
doubly-occupied sites $(1,1)$ and $(M,M)$ and focused on the
localisation length $\lambda_2$ obtained from the decay of the
transmission probability from one end of the system to the other,
i.e.,
\begin{equation}
  {1\over\lambda_2} = - {1\over\vert M-1\vert} \ln\vert\langle
  1,1\vert { G_2}\vert M,M\rangle\vert.
\label{eq-lambda2}
\end{equation}

%
%

\begin{figure}[t]
\centerline{\epsfig{figure=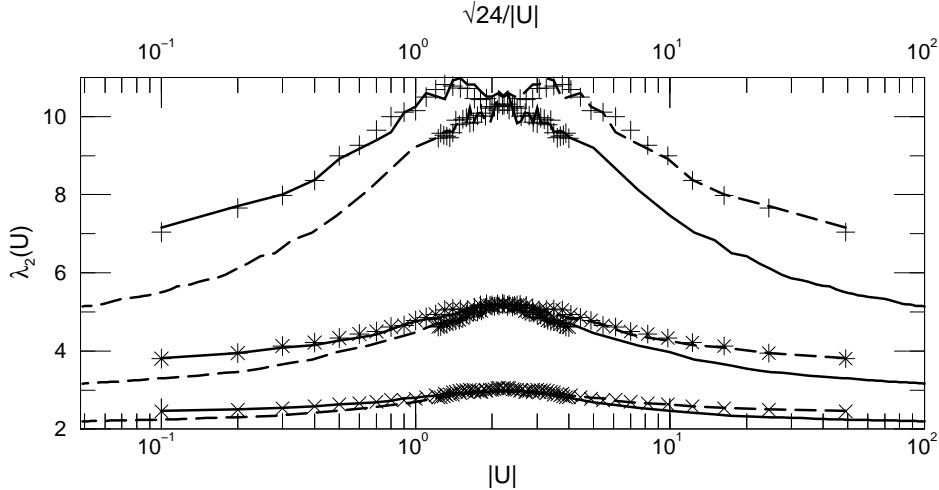,width=\figwidth} }
\caption{\label{fig-tipdm-l2_largeu}
%
  $\lambda_2(U)$ for TIP as a function of $|U|$ (solid lines) and
  $\protect\sqrt{24}/|U|$ (dashed lines) at $E=0$ for disorders $W=3$
  ($+$), $4$ ($*$), and $5$ ($\times$) and $M=201$.  The data are
  averaged over 100 samples. The lines (symbols) indicate data for
  $U>0$ ($U<0$).}
\end{figure}

In Fig.\ \ref{fig-tipdm-l2_largeu} we present data for $\lambda_2(U)$
obtained for three different disorders for system sizes $M=201$ at
$E=0$. 
In agreement with the previous arguments and calculations
\cite{RS97,PS97,WWP98}, we find that the enhancement is symmetric in
$U$ and decreases for large $|U|$. In \cite{WWP98} is has been argued
that at least for $\lambda_1 \approx M$, there exists a critical
$U_c={24}^{1/4}\approx 2.21$, which should be independent of $W$, at
which the enhancement is maximal.  We find that in the present case
with $\lambda_1 < M$ the maximum of $\lambda_2(U)$ depends somewhat on
the specific value of disorder used. The data in Fig.\ 
\ref{fig-tipdm-l2_largeu} may be compatible with the duality of Ref.\ 
\cite{WWP98}, but only for the large disorder $W=5$. For the smaller
disorders and for the range of interactions shown, we do not observe
the duality. We emphasize that the duality observed in \cite{WWP98} is
for spectral statistics and need not apply to quantities such as the
localisation length $\lambda_2$.

%
%

\begin{figure}[th]
  \centerline{\epsfig{figure=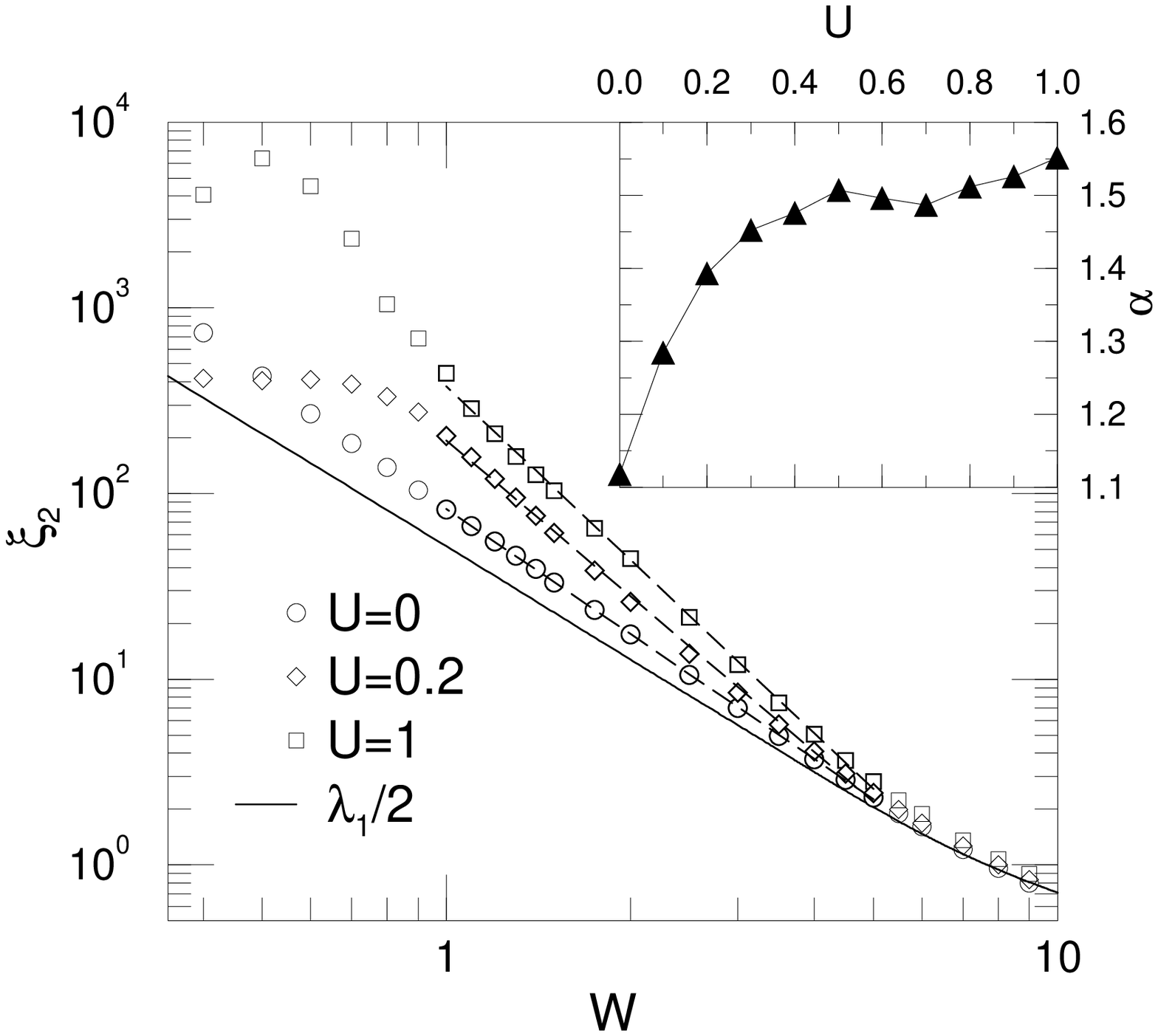,width=\figwidthb}
    \epsfig{figure=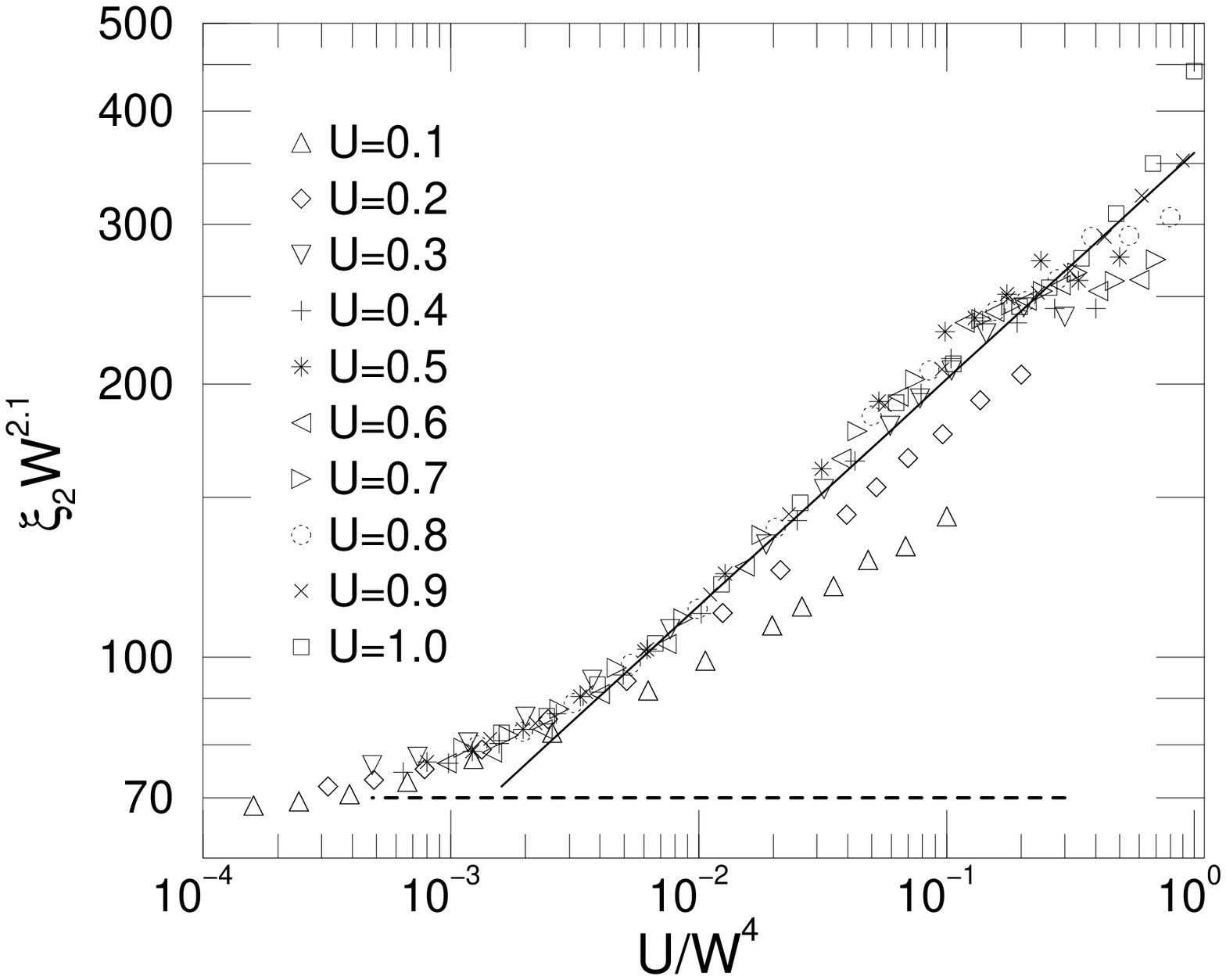,width=\figwidthb} }
\caption{\label{fig-tipdm-xi2_w}
%
  Left panel: TIP localisation lengths $\xi_2$ after FSS. The dashed
  lines represent power-law fits. Inset: Exponent $\alpha$ obtained by
  the power-law fits.
%
  Right panel: Scaling plot according to \protect\cite{SK97} with TIP
  localisation lengths $\xi_2(U)$ for $W\in [1,5]$. The solid line
  indicates a slope of 1/4, the dashed line the value of $\xi_2
  W^{2.1}$ in the limit $U=0$. }
\end{figure}

In order to reduce the possible influence of the finiteness of the
chain length, we constructed finite-size-scaling (FSS) curves for 11
interaction values $U= 0, 0.1, \ldots, 1$ from the $\lambda_2$ data
for $26$ disorder values $W$ between $0.5$ and $9$, for $24$ system
sizes $M$ between $51$ and $251$, averaging over 1000 samples in each
case.  In Fig.\ \ref{fig-tipdm-xi2_w} we show the infinite-size
localisation lengths (scaling parameters) $\xi_2$ obtained from these
11 FSS curves.
A simple power-law fit $\xi_2 \propto W^{-2\alpha}$ in the disorder
range $W\in [1,5]$ yields an exponent $\alpha$ which increases with
increasing $U$ as shown in the inset of Fig.\ \ref{fig-tipdm-xi2_w},
e.g., $\alpha= 1.55$ for $U=1$ and $\alpha= 1.1$ for $U=0$.  Because
of the latter, in the following we will compare $\xi_2(U\neq 0)$ with
$\xi_2(0)$ when trying to identify an enhancement of the localisation
lengths due to interaction.

Song and Kim \cite{SK97} suggested that the TIP localisation data may
be described by a scaling form $\xi_2 = W^{-\alpha_0}
g(|U|/W^{\Delta})$ with $g$ a scaling function. They obtain $\Delta =
4$ by fitting the data. Our data can be best described when $\alpha_0$
is related to the disorder dependence of $\xi_2$ as $(\alpha -
\alpha_0)/\Delta \approx 1/4$. As shown in Fig.\ 
\ref{fig-tipdm-xi2_w}, the scaling is only good for $W\in [1,5]$ and
$U \geq 0.3$. We note that assuming an interaction dependent exponent
$\alpha(U)$, we still do not obtain a good fit to the scaling function
with the data for all $U$.

\begin{figure}[th]
  \centerline{\epsfig{figure=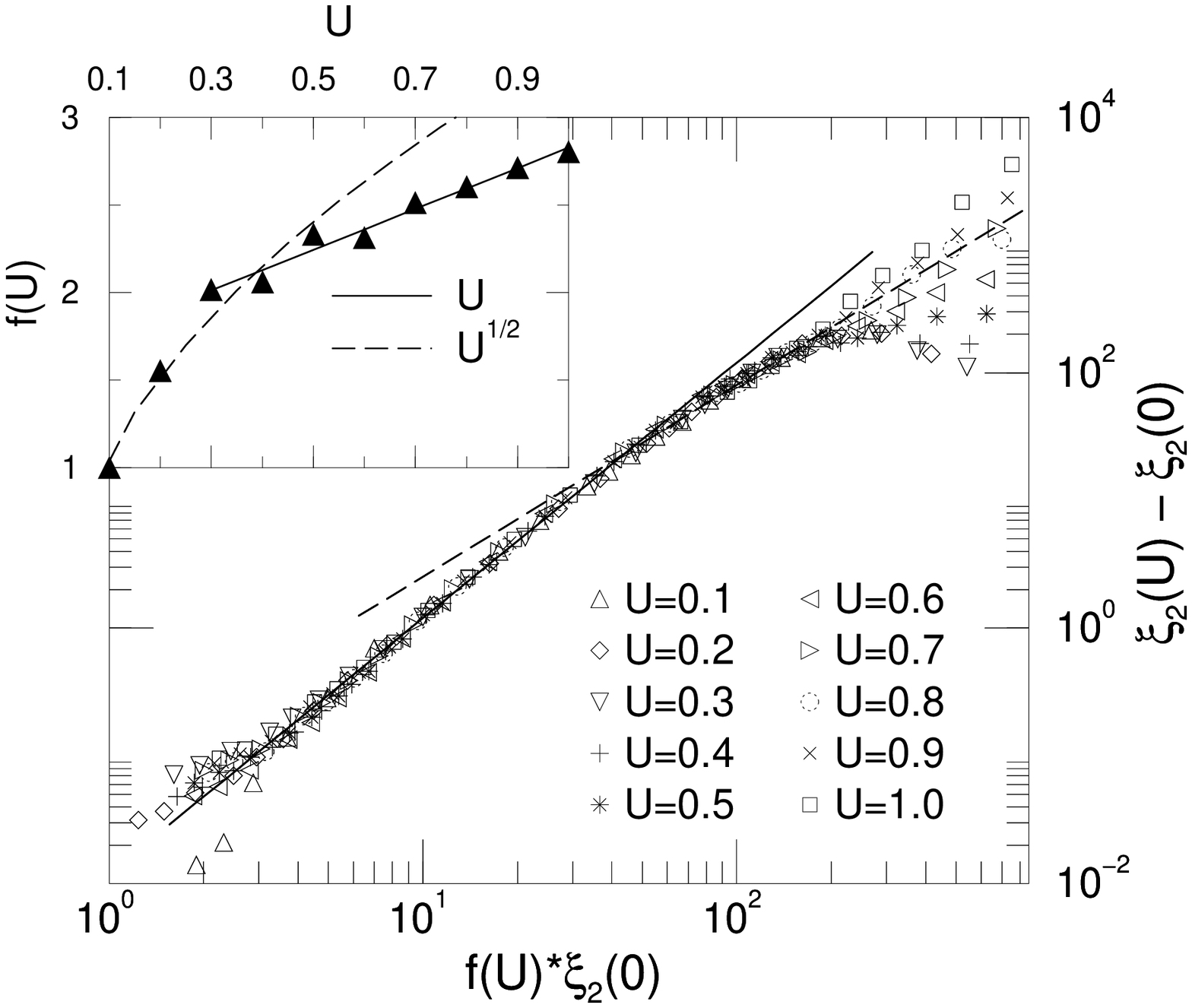,width=\figwidthb}
    \epsfig{figure=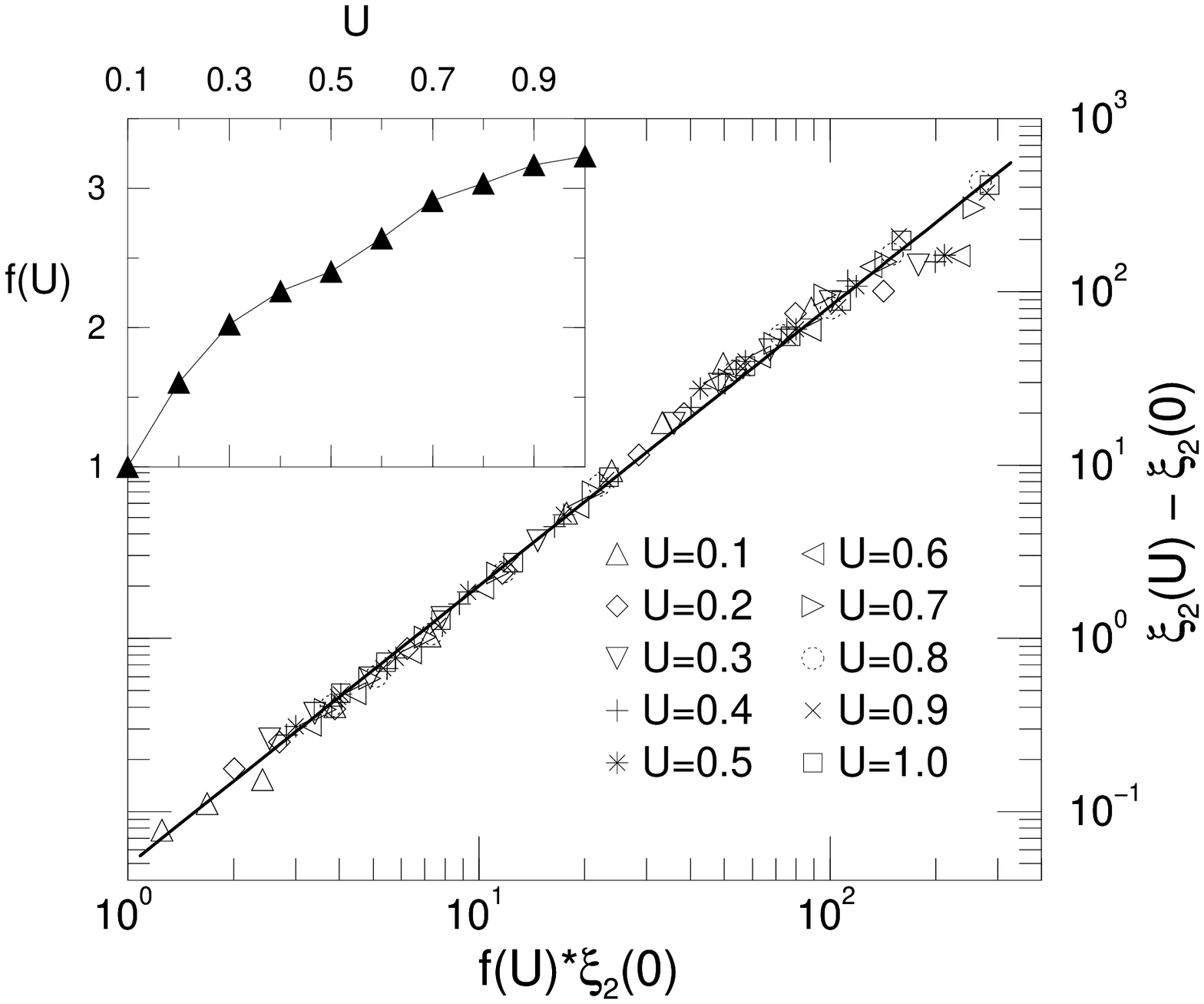,width=\figwidthb}}
\caption{\label{fig-tipdm-vo}
%
  Left panel: Scaling plot of Eq.\ (\protect\ref{eq-vo}) for TIP with
  $W\in [0.6,9]$. The solid (broken) line indicates the slope 2 (1.5).
  Inset: The values of $f(U)$ needed to make the data collapse onto
  the $U=0.1$ curve. 
%
  Right panel: IEH localisation length scaled as in the left panel.
  The solid line indicates slope 1.61 and $W\in [1,7]$.}
\end{figure}

In Fig.\ \ref{fig-tipdm-vo}, we show that a much better scaling can be
obtained when plotting
\begin{equation}
  \xi_2(U) - \xi_2(0) = \tilde{g}\left[ f(U) \xi_2(0) \right]
\label{eq-vo}
\end{equation}
with $f(U)$ determined by FSS. Now the scaling is valid for {\em all}
$U$ and $W\in [0.6,9]$. As indicated by the straight lines, we observe
a crossover from a slope $2$ to a slope $3/2$. There are some
deviations from scaling, but these occur for large and very small
values of $\xi_2(U)$ and are most likely due to numerical inaccuracy
\cite{LRS98a}. In the inset of Fig.\ \ref{fig-tipdm-vo}, we show the
behavior of $f(U)$. For $U \geq 0.3$ a linear behavior $f(U) \propto
U$ appears to be valid which translates into a $U^2$ ($U^{3/2}$)
dependence of $\xi_2(U) - \xi_2(0)$ in the regions of Fig.\ 
\ref{fig-tipdm-vo} with slope $2$ ($3/2$). For $U \leq 0.5$, we have
$f(U)\propto \sqrt{U}$ which yields $\xi_2(U) - \xi_2(0) \propto U$
($U^{3/4}$).

Thus in summary it appears that our data cannot be described by a
simple power-law behavior with a single exponent as in Eq.\ 
(\ref{eq-shep}) neither as function of $W$, nor as function of
$\xi_2(0)$ \cite{LRS98a}, nor after scaling the data onto a single
scaling curve.

%
%


As for TIP we computed \cite{LRS98b} the IEH localisation lengths by
the DM along the diagonal using $100$ realizations for each $(U,M,W)$.
We find that the data for IEH are very similar to the case of TIP. We
again perform FSS and observe that the infinite-size estimates
$\xi_2(U)$ are well characterized by an exponent $\alpha(U)$. We can
again scale the $\xi(U)$ data for IEH onto a single curve as shown in
Fig.\ \ref{fig-tipdm-vo}. However, here the crossover from slope $2$
to $3/2$ is much less prominent and the data can be described
reasonably well by a single slope of $1.61$. Also, the crossover
behavior in $f(U)$ is suppressed. We remark that these differences may
be due to the smaller number of samples used for IEH.

%
%


In conclusion, we observe an enhancement of the two-particle
localisation length due to onsite interaction both for TIP and IEH.
This enhancement persists, unlike for TMM \cite{RS97,RS98,HMK98}, in
the limit of large system size and after constructing
infinite-sample-size estimates from the FSS curves. We remark that the
IEH case is of relevance for a proposed experimental test of the TIP
effect \cite{G98}.

\ack R.A.R. gratefully acknowledges support by the Deutsche
Forschungsgemeinschaft (SFB 393).

%
%


\begin{thebibliography}{99}
\frenchspacing




\bibitem{LRS98a} M. Leadbeater, R. A. R\"{o}mer, and M. Schreiber,
submitted to Eur. Phys. J. B, (1998, cond-mat/9806255).

\bibitem{LRS98b} M. Leadbeater, R. A. R\"{o}mer, and M. Schreiber,
submitted to Phys. Rev. B, (1998, cond-mat/9806350).

\bibitem{S94} D. L. Shepelyansky, Phys. Rev. Lett. {\bf 73}, 2607
  (1994); F. Borgonovi and D. L. Shepelyansky, Nonlinearity {\bf 8},
  877 (1995); ---, J. Phys. I France {\bf 6}, 287 (1996); Y. Imry,
  Europhys. Lett. {\bf 30}, 405 (1995).




\bibitem{FMPW95} K. Frahm, A. M\"uller-Groeling, J.-L. Pichard, and D.
  Weinmann, Europhys. Lett. {\bf 31}, 169 (1995);
%
D. Weinmann, A. M\"{u}ller-Groeling, J.-L. Pichard, and
  K. Frahm, Phys. Rev. Lett. {\bf 75}, 1598 (1995);
%
F. v. Oppen, T. Wettig, and J. M\"uller, Phys. Rev.
  Lett.  {\bf 76}, 491 (1996);
%
T. Vojta, R. A. R\"omer, and M. Schreiber, preprint
  (1997, cond-mat/9702241);
%
  D. Brinkmann, J. E. Golub, S. W. Koch, P. Thomas, K. Maschke, and I.
  Varga, preprint (1998).

\bibitem{SK97} P. H. Song and D. Kim, Phys. Rev. B {\bf 56}, 12217
  (1997).

\bibitem{RS97} R.  A. R\"omer and M. Schreiber, Phys. Rev.  Lett. {\bf
    78}, 515 (1997); K. Frahm, A. M\"{u}ller-Groeling, J.-L.  Pichard,
  and D. Weinmann, Phys. Rev. Lett. {\bf 78}, 4889 (1997); R.  A.
  R\"omer and M. Schreiber, Phys. Rev.  Lett. {\bf 78}, 4890 (1997).

\bibitem{PS97} I. V. Ponomarev and P. G. Silvestrov, Phys. Rev. B {\bf
    56}, 3742 (1997).

\bibitem{RS98} R.  A. R\"omer and M. Schreiber, phys.  stat.  sol. (b)
  {\bf 205}, 275 (1998).

\bibitem{HMK98} O. Halfpap, A.  MacKinnon, and B. Kramer, Sol. State
  Comm., (1998), in print.

\bibitem{LW80} C. J. Lambert and D. Weaire, phys. stat. sol. (b) {\bf 101},
  591 (1980).


\bibitem{WWP98} X. Waintal, D. Weinmann, and J.-L. Pichard, preprint
  (1998, cond-mat/9801134).



\bibitem{G98} J. E. Golub, private communication.








\end{thebibliography}
\end{document}